\documentclass[runningheads]{llncs}
\usepackage[T1]{fontenc}
\usepackage{graphicx}
\usepackage{xcolor}
\usepackage{multirow}
\usepackage{array}
\usepackage{subfigure}
\usepackage{subcaption}
\usepackage{lineno}
\usepackage{xstring}
\usepackage{hyperref, cleveref}
\usepackage{enumitem}
\usepackage{todonotes}
\newif\ifblind
 
\makeatletter
\IfSubStr{,\@classoptionslist,}{,review,}{%
  \blindtrue   % Found it!
}{%
  \blindfalse  % Not found
}
\makeatother
 
\newcommand{\anon}[2][Anonymised for Review]{%
  \ifblind%
    \textcolor{blue}{[#1]}%
  \else%
    #2%
  \fi%
}
 
\ifblind
  \linenumbers
\fi

\begin{document}
\title{
% "How Do I Do This?"
% ``How Do I \dots?'': Procedural Questions Predominate Student-LLM Chatbot Interactions
``How Do I \dots?'': Procedural Questions Predominate Student-LLM Chatbot Conversations
% ``How Do I \dots?'': Procedural Questions Predominate Conversations between Students and LLM Chatbots
% ``How Do I \dots?'': Students Predominantly Ask LLM Chatbots Procedural Questions
% ``How Do I \dots?'': What Student Questions Reveal About LLM Chatbot Use While Learning?
% ``How Do I \dots?'': What Student Questions Reveal About LLM Chatbot Use in Learning Contexts?
}
\titlerunning{``How Do I \dots?'': Procedural Questions Predominate \dots}
\author{
    \anon[Anonymised Author(s)]{
    Alexandra Neagu\orcidID{0009-0004-0840-3776} \and
    Marcus Messer\orcidID{0000-0001-5915-9153} \and
    Peter Johnson\orcidID{0000-0001-7841-691X} \and
    Rhodri Nelson\orcidID{0000-0003-2768-5735}
    }
}
\authorrunning{\anon[Anonymised Author(s)]{A. Neagu et al.}}
\institute{
\anon[Anonymised Institution]{Imperial College London, London, United Kingdom}
\email{\anon[Anonymised Email]{alexandra.neagu20@imperial.ac.uk}}\\
}
\maketitle              % typeset the header of the contribution
\begin{abstract}
% The abstract should briefly summarize the contents of the paper in 150--250 words.
% 150 to 250 words. The abstract should not contain any undefined abbreviations or unspecified references.
% The first sentence should be something familiar to all readers, to set the context.
% Background:

% ----Alex2nd Feb
Providing scaffolding through educational chatbots built on Large Language Models (LLM) has potential risks and benefits that remain an open area of research. When students navigate impasses, they ask for help by formulating impasse-driven questions. Within interactions with LLM chatbots, such questions shape the user prompts and drive the pedagogical effectiveness of the chatbot's response. 
This paper focuses on such student questions from two datasets of distinct learning contexts: formative self-study, and summative assessed coursework. We analysed 6,113 messages from both learning contexts, using 11 different LLMs and three human raters to classify student questions using four existing schemas. 
On the feasibility of using LLMs as raters, results showed moderate-to-good inter-rater reliability, with higher consistency than human raters. 
The data showed that `procedural' questions predominated in both learning contexts, but more so when students prepare for summative assessment. 
These results provide a basis on which to use LLMs for classification of student questions.
However, we identify clear limitations in both the ability to classify with schemas and the value of doing so: schemas are limited and thus struggle to accommodate the semantic richness of composite prompts, offering only partial understanding the wider risks and benefits of chatbot integration. 
In the future, we recommend an analysis approach that captures the nuanced, multi-turn nature of conversation, for example, by applying methods from conversation analysis in discursive psychology.

% Inter-Rater Reliability
\end{abstract}

% ----- Sections ----
\section{Introduction}

% new plan (as in the new abstract):
% - Learning involves impasse and to overcome impasse students need scaffolding
% - LLM chatbots can provide scaffolding, but they are still a disruptive tech with risks and benefits to keep in mind
% - also chatbot effectiveness depends on their inputs and that involves the student's use pateerns and their questions
% - this paper looks into the student questions and what insight we can get from them in the use patterns of chatbots in two different learning context: formative summative
% - one way to analyse question has been applying classification schemas, which has been done in educational literature in tutoring and ITS fields
% - such schemas categorised student quedtions in ...examples of classes...
% - but little work has been implemented on student-chatbot conversations directly
% - especially applyig the schemas at scale, as such analysis is very time consuming on the raters, this paper aims to explore the feasability of scaling it up by using LLMs as raters.
% - So our RQ are:...
% - we did this by METHODOLOGY - detail
% - implications and contributions

Learning is a stateful subjective process of navigating struggles, including moments of impasse \cite{VanLehn1988impasse}. 
For students to progress and overcome impasses they cannot resolve alone, they require appropriate scaffolding from an external source \cite{Vygotsky1978ZPD}. 
Educational chatbots driven by Large Language Models (LLMs) can serve as this scaffolded intervention during learning \cite{labadze2023ChatbotsInEdu,neagu2025LLMsFeedback}. 
While these chatbots constitute an opportunity, such as increased efficiency \cite{CasebourneWegerif2024,jurenka2024responsibledevelopmentgenerativeai,labadze2023ChatbotsInEdu}, they also carry risks such as undesirable cognitive offloading \cite{neagu2025LLMsFeedback,wollny2021chatbotsInEduReview}). Research is required to understand the risks and benefits of chatbots. 

The effectiveness of LLM chatbots during learning depends on its inputs, which are driven by the user prompts, and in the case of educational chatbots, student prompts \cite{ammari2025students}. 
Therefore, research into chatbot effectiveness requires an understanding of student prompts, which often take the form of questions. 
However, educational literature indicates students often struggle to verbalise their needs through prompts to LLM chatbots \cite{Ma2025promptingHowTo,zamfirescu2023nonexpertPromptingChal}, mirroring difficulties in articulating questions observed in traditional human-driven and Intelligent Tutoring Systems (ITS) tutoring \cite{anthony2004student,graesser1994question}. 
Therefore, to gain insights into how students prompt LLM chatbots, this paper analyses student questions from two distinct STEM (Science, Technology, Engineering, and Mathematics) learning contexts: formative problem-solving self-study in engineering, and summative coursework task in computing.

Schemas can be used to classify student questions and have been developed and applied in the fields of human tutoring and Intelligent Tutoring Systems (ITS) \cite{anthony2004student,caowang2021qgen,graesser1994question,harrak2019student}. 
For example, under such schemas question can be classified as `procedural' (\textit{``how do we derive it''}), `causal' (\textit{``why is pressure not a function of theta and z''}), `definition' (\textit{``what are the units for viscosity''}), or `verification' (\textit{``do we assume the pressure gradient is 0?''}). 

Application of classification schemas onto student prompts to LLM-driven chatbots is an open area of research \cite{ammari2025students,koyuturk2025learnerchatbotInteractions,maiti2024studentsinteractllmpoweredvirtual}. However, applying schemas at scale is time-consuming for human raters \cite{bavaresco2025llmsAsJudge}, hence this paper focuses on evaluating the feasibility of using LLM automation to analyse datasets too large for human classification.
Accordingly, this paper explores the following research questions:
\begin{enumerate}[label={\textbf{RQ\arabic*}}, align=left]
    \item How feasible can existing classification schemas be used to analyse student questions to LLM chatbots?
    \item What kind of questions do STEM students ask LLM chatbots during a formative problem-solving task versus a summative coursework task?
\end{enumerate}

% To address these questions, this paper analysed two student-chatbot conversation datasets from the STEM field. 
\Cref{section:related_work} details the four question classification schemas used to classify a total of 6,113 student messages by 11 LLM and three human raters, following the process detailed in \Cref{section:method}.
% across four schemas. 
Finally, \Cref{results} and \Cref{discussion} present the insights and contributions of this work: a robust methodology for using LLM-raters for question classification; and an in-depth analysis into what types of questions students ask LLM chatbots in two contrasting learning contexts.

\section{Background}\label{section:related_work}

This section introduces the four student-focused question classification schemas from ITS, flipped classroom activities, and open-ended student questions, applied by the LLM raters described in \Cref{section:method}, and weighs the balance between automated classification approaches and the costly human gold standard.

\subsection{Question Classification Schemas}\label{rw:schemas}

% To evaluate LLM chatbots in education, research has analysed the context in which the interaction occurs \cite{winklersoellner2018ChatbotsinEduContext}. One method is to apply schemas to classify the kind of questions students ask \cite{anthony2004student,graesser1994question,harrak2019student}.

Schemas can identify cognitive mechanisms behind questions, such as a student detecting a knowledge gap or monitoring for shared understanding \cite{graesser1994question}. These insights inform pedagogical improvements \cite{graesser1994question} and computational applications \cite{anthony2004student}. 
For instance, Graesser and Person \cite{graesser1994question} used schemas to identify that students in human tutoring mostly ask low-level questions, highlighting the need for explicit training in effective questioning to improve comprehension, learning, and memory.
Similarly, Anthony et al. \cite{anthony2004student} found that users of ITS asked more questions when unprompted, suggesting a need for better scaffolding within the ITS to encourage self-reflection. 

While schemas are increasingly applied to student-questions to LLM chatbots \cite{ammari2025students,koyuturk2025learnerchatbotInteractions,maiti2024studentsinteractllmpoweredvirtual}, existing studies typically focus on specific, constrained learning tasks (e.g., coding \cite{mcnichols2025studychat}) or lack a comparative analysis with multiple classification frameworks and methods.
% martinenghi2024VonneumidasSchema
Therefore, this paper used the following four question classification schemas as they originate from different research fields, aim to be independent of an educational task, and focus on the question semantics:

\textbf{\textit{Graesser1994}} \cite{graesser1994question} (18 classes) was originally designed to analyse one-to-one human tutoring sessions. It classifies questions based on content, psychological mechanism, and specification. It distinguishes between ``shallow'' questions (demanding short answers, e.g., \textit{`verification'}) and ``deep'' questions (demanding long answers, e.g., \textit{`causal'} or \textit{`procedural'}).

\textbf{\textit{Anthony2004}} \cite{anthony2004student} (5 classes) was developed via a study simulating a conversational ITS for mathematics. It categorises questions by depth --- \textit{`answer-'}, \textit{`process-'}, or \textit{`principle-oriented'} --- and notably included an \textit{`interface'} category (e.g., how to use the software). While useful in its original context, the \textit{`interface'} category might be challenging when applied to data from different learning contexts as the term can refer outside of the task or chatbot software.

\textbf{\textit{Harrak2019}} \cite{harrak2019student} (14 classes) focused on a flipped-classroom setting for medical students. Unlike the other schemas, this schema uses a bottom-up, data-driven approach based on keywords to identify student intent. For example, words like ``explain'' classify a question as \textit{`deepen a concept'}, while ``can we'' or ``is it'' classifies it as \textit{`validation'}.

\textbf{\textit{Cao2021}} \cite{caowang2021qgen} (10 classes) focused on generating open-ended questions that require deep comprehension. Although not strictly an educational study, their work aimed to capture deep cognitive levels, such as causal reasoning and judgement. They adapted their schema from Graesser and Person’s earlier work \cite{graesser1994question,olney2012question} by annotating data from active online forums, specifically Yahoo Answers and scientific and historical Reddit communities.

\subsection{Scalable Classification}

Human labelling is currently considered the gold standard for nuanced, subjective data, such as student-chatbot conversations, but its high cost and scalability are limiting factors \cite{zheng2023judgingLLMjudge}. These limitations motivate automated classification.

Specialised machine learning models, such as fine-tuned BERT, can perform well-defined classification tasks, such as detecting ``talk moves'' in classroom discourse \cite{wang2023bertLLMtalkMoves}. However, these models require extensive labelled training data.

Large Language Models (LLMs) can perform classification on unstructured inputs without further specialised training \cite{zheng2023judgingLLMjudge}. Those LLM capabilities are particularly valuable in data-scarce environments where extensive training data is unavailable \cite{alfaraby2024BertLLMsQclass,ammari2025students}. LLMs are susceptible to training biases, including preferences for longer, more assertive, or outputs similar to its own \cite{zheng2023judgingLLMjudge}. 

There is a trade-off between human, fine-tuned models, and LLM-raters in classification tasks \cite{gani2025tfidfLLMsQclass,wang2023bertLLMtalkMoves,zheng2023judgingLLMjudge}. Therefore, a balanced approach is required where LLMs augment, rather than entirely replace, human oversight.

% For example, one study demonstrated that zero-shot prompting of GPT-3.5-turbo achieved a higher F1-score than a trained DistillBERT + Random Forest model, highlighting the potential of LLMs even without large training sets \cite{alfaraby2024BertLLMsQclass}.

\section{Method}\label{section:method}

% To analyse what questions students ask LLM chatbots within the two datasets detailed in \Cref{m:datasets}, we classified student questions at scale using LLM-raters. The process was structured into two pipelines: \Cref{m:processingPipeline} details how student messages were filtered for questions, and classified using four schemas. \Cref{m:evaluationPipeline} validates the robustness of the labels by conducting inter-rater reliability analysis across 11 LLM-raters and verifying validity against three human raters.

% Two datasets of student-chatbot conversations were used for analysis and are described in \Cref{m:datasets}. The data was processed in two stages: first filtering conversation messages for student questions, described in \Cref{m:filtering}; and second classifying the questions using schemas, described in \Cref{m:classification}. The classification used 11 LLM and three human raters, allowing for evaluation tests by conducting inter-rater reliability analysis, described in \Cref{m:evaluationPipeline}.

% \begin{figure}
%     \centering
%     \includegraphics[width=1\linewidth]{misc/QuestionAnalysisPipeline-MethodPipeline_v2.drawio.pdf}
%     \caption{Method of analysing student questions within conversations with chatbots}
%     \label{fig:methodPipeline}
% \end{figure}

\subsection{Datasets of Two Learning Contexts} \label{m:datasets}

Two datasets of student–LLM chatbot conversations were used for analysis.
% : one collected from an internal learning platform and one publicly available dataset. 
The primary distinction between the datasets was their learning context --- a formative self-study setting in engineering, versus a summative coursework setting in computer science.
\Cref{tab:exampleDatasets} shows conversations snippets from the two datasets.

The \textbf{\texttt{FormativeChat}} dataset (internal) consists of student-chatbot conversations gathered over two months from \anon{Lambda Feedback \cite{johnson2025lambdafeedback}}, an interactive self-study platform deployed at a UK university. The dataset covers conversations from 41 second year undergraduate students that interacted with an LLM chatbot on formative tasks (but relevant to summative assessments) for a Engineering course on Fluid Mechanics. The chatbot, powered by gemini-2.0-flash, was equipped with learning context --- a system prompt with information about the tasks the students were working on and their progress on those tasks. 
The students had access to task solutions regardless of their chatbot use.
% Students were notified of the data usage for research purposes via the platform’s privacy policy.
Ethics approval to use the data for research purposes was granted (\anon{EERP2526-059}).

The \textbf{\texttt{SummativeChat}} dataset (public, originally StudyChat \cite{mcnichols2025studychat}) captures a summative learning context. It contains anonymised student-LLM chatbot conversations collected during a semester-long Computer Science course on AI at a US university. In the study, 203 students used the chatbot to get support with Python programming assignments that contributed directly to their final grade. Students were encouraged to use the chatbot freely, with their use-patterns having no academic penalty. The chatbot, built on gpt-4o-mini, lacked access to specific course materials as it functioned similarly to ChatGPT. 
% The original study reported students primarily seeking practical coding assistance and conceptual explanations.

\begin{table}[ht]
    \centering
    \renewcommand{\arraystretch}{1.2}
    \setlength{\tabcolsep}{0.4em}
    \footnotesize
    \caption{Conversation snippets from the two datasets reflecting a formative self-study (\texttt{FormativeChat}) and summative assessment setting (\texttt{SummativeChat})}
    \begin{tabular}{|p{0.48\linewidth} | p{0.48\linewidth}|} 
        \hline
        \centering\textbf{\texttt{FormativeChat}} & \centering\textbf{\texttt{SummativeChat}} \arraybackslash \\
        \hline
        \textbf{Student:} \textit{``How can I start solving this question?''} \newline 
        \textbf{Chatbot:} \textit{``Okay, let's get you started on Part (a) of the racing car \dots \newline
        % question. The key to this part is understanding the definition of bulk velocity \dots \newline \newline
        \dots [proposes 3 steps to focus on] \dots \newline
        If you get stuck, let me know where and I can offer more specific guidance.''} \newline
        \textbf{Student:}\textit{``lets go [through] it step by step''}
        \newline\parbox{\linewidth}{\centering \textit{\dots conversation continues}\dots}
        & 
        \textbf{Student:} \textit{``[section of student code] Why does this only print the second one''} \newline 
        \textbf{Chatbot:} \textit{``If you're running this code in an interactive environment \dots\ you might see 
        % output only for the last expression evaluated in a cell. \newline
        \dots [proposes 3 code fixes] \dots \newline
        Using any of the above methods should show both datasets.''}\newline
        \textbf{Student:} \textit{``For all loop in python''}
        \newline\parbox{\linewidth}{\centering \textit{\dots conversation continues\dots}}
        \\
        \hline
    \end{tabular}
    \label{tab:exampleDatasets}
\end{table}

% \subsection{Processing Pipeline}\label{m:processingPipeline}

% \noindent\textbf{Filtering: } 
\subsection{Conversation Filtering} \label{m:filtering}
Prior to the classification of student questions, a two-stage pre-filtering pipeline was applied on the datasets to standardise the question sets, ensuring consistency across the classification tasks for all schemas.
Firstly, potential questions were identified using symbolic detection through regular expression filters, including matching w-words (\textit{who, what, when, where, why}), question marks, or phrases such as \textit{``can I...''}. Any remaining messages that were not classified as questions were then re-evaluated by an LLM (gemini-2.0-flash). The LLM was prompted with Graesser and Person's definition of a `question' \cite{graesser1994question}, describing it as an inquiry, an interrogative expression, or both.
For example, \textit{``explain variability in easy terms''} was missed by symbolic filters but correctly detected as a `question' by the LLM.

\subsection{Question Classification}\label{m:classification}\label{m:schemas}
% \noindent\textbf{Classification: } 
% \subsubsection{Classification: } 
The four question classification schemas detailed in \Cref{rw:schemas} (\textit{Graesser1994}~\cite{graesser1994question}, \textit{Anthony2004}~\cite{anthony2004student}, \textit{Harrak2019}~\cite{harrak2019student}, \textit{Cao2021}~\cite{caowang2021qgen}) were applied on the student-chatbot conversations of the two datasets from \Cref{m:datasets}.
Classification was done by 11 LLMs of varying sizes, sources and providers from \Cref{tab:LLM-configurations}. Each question was provided to the LLM-raters along with a prompt describing one of the classification schemas. The prompt included the classification task description and all schema details such as class labels, definitions, examples, or keywords.

\begin{table}[h]
    \renewcommand{\arraystretch}{1.3}
    \setlength{\tabcolsep}{0.3em}
    % \footnotesize
    \centering
    \caption{LLM Models Used as Raters for Question Classification}
    \begin{tabular}{|>{\centering}p{0.16\linewidth}|>{\centering}p{0.19\linewidth}|p{0.6\linewidth}|}
        \hline
        \multirow{3}{*}{\parbox{\linewidth}{Open-source models}} 
            & Small / Medium& phi4:14b-q8\_0, gpt-oss:20b, gpt-oss:120b \\
            \cline{2-3}
            & Large & llama4:128x17b \\
            \cline{2-3}
            & Reasoning & deepseeks-r1:70b \\
        \hline
        \multirow{4}{*}{\parbox{\linewidth}{Closed-source models}} 
            & \multirow{2}{*}{Small / Medium} & gpt-4o-mini(2024-07), gpt-5-nano(2025-08)\\
            & &  gemini-2.0-flash(2025-02), gemini-2.5-flash(2025-06) \\
            \cline{2-3}
            & Large & gpt-4.1(2025-04)\\
            \cline{2-3}
            & Reasoning & gpt-5.1(2025-11, reasoning effort: medium) \\
        \hline
    \end{tabular}
    \label{tab:LLM-configurations}
\end{table}

\subsection{Evaluation of Question Classification}\label{m:evaluationPipeline}

To evaluate the reliability and validity of using LLMs for categorising student questions, we compared the outputs of 11 LLM-raters (\Cref{tab:LLM-configurations}) and three human raters across two datasets (\Cref{m:datasets}) and four classification schemas (\Cref{m:schemas}). The evaluation followed four steps: testing inter-rater agreement, evaluating agreement variance between raters, verifying internal consistency of LLM-raters, and validating LLM against human labels.

Inter-rater reliability used the Fleiss’ Kappa \cite{fleiss1971kappa} coefficient, and was calculated due to having more than two raters. Datasets of subjective nature may exhibit high prevalence in certain categories, which means Fleiss’ Kappa can be artificially lowered by uneven distributions \cite{derksen2024kappaparadox}. To account for this statistical paradox, we also calculated Gwet's AC1 \cite{Gwet2008}, which provides a more robust measure of agreement in the presence of high-prevalence categories. This analysis ensured the labelling process was not biased by class imbalances.

Secondly, we assessed variability among the raters through a leave-one-out analysis \cite{gwet2014handbook}. The process involved iteratively calculating inter-rater reliability coefficients --- Fleiss’ Kappa and Gwet's AC1 --- while removing one rater from the pool. An increase in the reliability coefficient following a rater’s removal from the pool identified that rater as a source of disagreement, while a decrease indicated that the rater contributed positively to the inter-rater agreement.

Thirdly, we analysed the internal consistency of the LLM-raters by evaluating their sensitivity to prompt formatting. We shuffled the order of class definitions within the system prompts of three LLM-raters (gemini-2.0-flash, gpt-4o-mini, gpt-oss:20b) across all four classification schemas and applied them on \texttt{FormativeChat}. The analysis involved comparing the two versions of labels from the same rater (original vs. randomised order) using Cohen's Kappa \cite{cohen1960kappa} as the inter-rater reliability coefficient. We then performed a Wilcoxon Signed-Rank Test to determine whether the presentation order of the schemas within the LLM system prompts significantly impacted the labelling process.

% \subsubsection{Human Validation:} 
% \noindent\textbf{Human Validation: }\label{m:human_labelling}
Finally, while inter-rater reliability coefficients measure consistency among raters, they do not indicate the validity of labels. 
To validate the correctness of the LLM-generated outputs, three human raters labelled a randomised subset of the \texttt{FormativeChat} dataset (10\% of the dataset) to establish a `ground truth' for comparison. These raters were familiar with the learning platform and its pedagogical context, ensuring that their labels reflected meaningful and contextually accurate interpretations of student questions.
We applied the same Leave-One-Out analysis as above to the mixed LLM and human-raters pool. 
By iteratively calculating inter-rater reliability coefficients with human raters included in the pool, we assessed how closely the LLM-raters aligned with human judgment.

% seed 42,
% NOTE: in results show that the random sample has a similar distribution as the overall data (correlation measure between the two)
\section{Results}\label{results}

Filtering student messages for questions resulted in a sample size of 943 student questions (70\% of messages) from \texttt{FormativeChat} and 5170 (80\% of messages) from \texttt{SummativeChat}.
\Cref{r:agreements} details the results of the inter-rater reliability analysis used to evaluate the LLM and human raters.
\Cref{results:proceduralQs} examines the specific types of questions students asked across the two datasets.
Supplementary material including more in-depth analysis and prompts can be found on OSF
% \footnote{\url{https://osf.io/uv42q/overview?view_only=bf1b707888204f97933db70e9262dac4}}
\footnote{Supplementary materials: \url{https://doi.org/10.17605/OSF.IO/UV42Q}}
.
% \cite{neagu2026howdoI}

\subsection{(RQ1) Question Classification Reliability} \label{r:agreements}\label{r:anthony_misalignment}

% \noindent\textbf{Agreement between LLM-Raters: } 
\subsubsection{Agreement between LLM-Raters: } 
\begin{figure}[]
    \centering
    \includegraphics[width=1\linewidth]{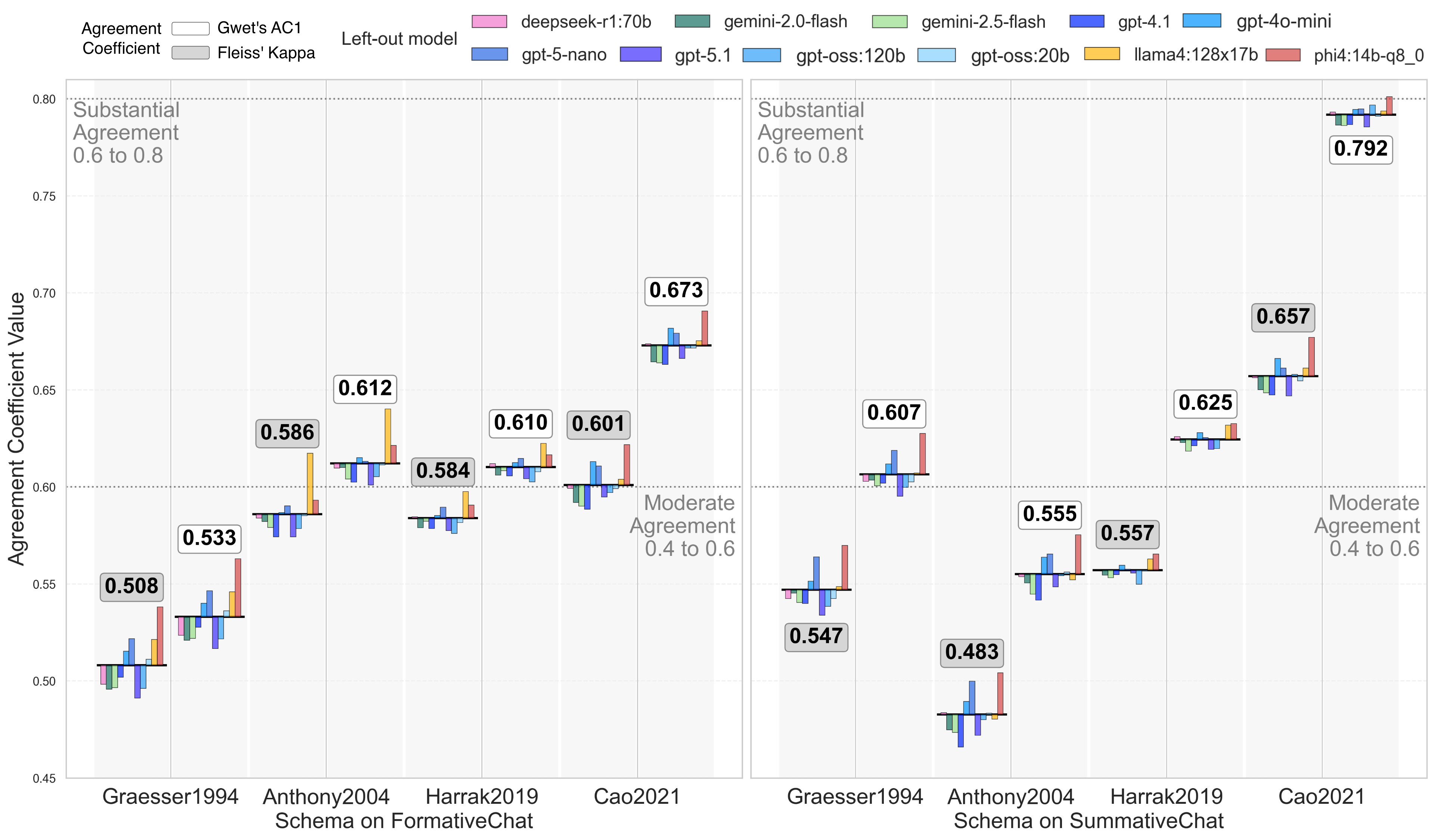}
    % \begin{subfigure}
    %     \centering
    %     \includegraphics[width=\linewidth]{misc/QuestionAnalysisPipeline-LLM_legend.drawio.png}
    % \end{subfigure}
    % \begin{subfigure}
    %     \centering
    %     \includegraphics[trim={0.2cm 0 5cm 0},clip, width=0.49\linewidth]{misc/aggregated_LOO_bar_FormativeChat_all_coeffs.png}
    % \end{subfigure}
    % \begin{subfigure}
    %     \centering
    %     \includegraphics[trim={0.2cm 0 5cm 0},clip, width=0.49\linewidth]{misc/aggregated_LOO_bar_SummativeChat_all_coeffs.png}
    % \end{subfigure}
    \caption{Leave-One-Out analysis of inter-rater reliability across 11 LLM-raters for \texttt{FormativeChat} (left) and \texttt{SummativeChat} (right). Bars represent the change in inter-rater agreement coefficients 
    % (Fleiss' Kappa and Gwet's AC1) 
    when a specific rater is iteratively removed. An increase identifies the removed rater as a source of disagreement, while a decrease indicates the rater contributed positively to the agreement.}
    \label{fig:looAgreementComparison}
\end{figure}

Given the large number of independent LLM-raters (11), the reliability coefficients from \Cref{fig:looAgreementComparison} indicate an overall moderate-to-good level of agreement. The mean Fleiss’ Kappa ($\kappa$) was 0.570 for \texttt{FormativeChat} and 0.561 for \texttt{SummativeChat}, aligning with established benchmarks for moderate agreement $(0.41 \leq \kappa \leq 0.60)$ \cite{landis1977measurementFleiss}.
Gwet's AC1 values were slightly higher, with a mean of 0.607 for \texttt{FormativeChat} and 0.644 for \texttt{SummativeChat}, being interpreted as moderate to substantial agreement $(0.61 \leq \textit{AC1} \leq 0.8)$ \cite{landis1977measurementFleiss}, strengthened by their narrow variations (95\% confidence intervals) and significant p-values ($p < .001$) \cite{cicchetti2001methodologicalGwetAC1} across all schemas.
% Considering that Gwet's AC1 followed a similar trend to Fleiss' Kappa provides confidence that the results were not artificially lowered by the paradox of high category prevalence.
As Gwet's AC1 followed a similar trend to Fleiss' Kappa, we can infer that the uneven-distributions paradox did not artificially lower the results.
Higher inter-rater disagreement was often encountered where student questions were ambiguous or overlapped across multiple categories in a schema, such as questions in \Cref{tab:qualitative_examples}.
% was labelled by all LLM-raters as \textit{reason\_why} in \textit{Harrak2019}, but proved challenging for \textit{Graesser1994} and \textit{Cao2021}.
% While \textit{``do you use the dynamic pressure like for when we found the shear force? but multiply [it] by Cd instead''} (from \texttt{FormativeChat}) was labelled by the majority as \textit{instrumental\_procedural} under \textit{Graesser1994}, the message proved challenging for applying \textit{Harrak2019}. 
% The consensus was fractured: \textit{link\_between\_concept} was chosen by three raters, while three other classes (\textit{validation\_verification}, \textit{manner\_how}, \textit{deepen\_concept}) were evenly selected by two raters each.

\begin{table}[]
\centering
\renewcommand{\arraystretch}{1.1}
\setlength{\tabcolsep}{0.4em}
\caption{Examples of Student Questions Labelled by LLM and Human Raters}
\begin{tabular}{|>{\centering\arraybackslash}p{0.24\linewidth}|>{\centering\arraybackslash}p{0.17\linewidth}|>{\raggedright\arraybackslash}p{0.29\linewidth}|
>{\centering\arraybackslash}p{0.08\linewidth}|
>{\centering\arraybackslash}p{0.11\linewidth}|
}
\hline
  \textbf{Question}&\textbf{Schema}& \textbf{Label} & \textbf{LLMs} & \textbf{Humans}\\
\hline
\multirow{11}{*}{\parbox{\linewidth}{(a)\newline\newline``but why can't you use drag force to that shear stress then sub [substitute] that in'' \newline\newline\texttt{FormativeChat} }}
    & \multirow{4}{*}{\textit{Graesser1994}} & {expectational}   & 5 & 2 \\
    &                       & {causal\_antecedent}      & 4 & 1 \\
    &                       & {instrumental\_procedural}& 1 & - \\
    &                       & {assertion}               & 1 & - \\
\cline{2-5}
    & \multirow{2}{*}{\textit{Anthony2004}}& {principle\_oriented} & 8 & 2 \\
    &                       & {process\_oriented}       & 3 & 1 \\
\cline{2-5}
    & \textit{Harrak2019}   & {reason\_why}             & 11 & 3 \\
\cline{2-5}
    & \multirow{4}{*}{\textit{Cao2021}}& {procedural}        & 6 & 1 \\
    &                       & {cause}                   & 5 & - \\
    &                       & {concept}                 & - & 1 \\
    &                       & {judgmental}              & - & 1 \\
\hline
\multirow{6}{*}{\parbox{\linewidth}{(b)\newline\newline``parameter vs \newline variable'' \newline\newline\texttt{FormativeChat} }}
    & \textit{Graesser1994} & {comparison}              & 11 & 3 \\
\cline{2-5}
    & \textit{Anthony2004}  & {definition}              & 11 & 3 \\
\cline{2-5}
    & \multirow{2}{*}{\textit{Harrak2019}}& {link\_between\_concepts} & 10 & 2 \\
    &                       & {define}                  & 1 & 1 \\
\cline{2-5}
    & \multirow{2}{*}{\textit{Cao2021}} & {comparison}  & 8 & 3 \\
    &                       & {concept}                 & 3 & - \\
\hline
\end{tabular}
\label{tab:qualitative_examples}
\end{table}

The leave-one-out analysis identified primary contributors to the inter-rater agreement as mainly larger models such as gpt-5.1 and gpt-4.1, with contributions also from smaller models such as the reasoning model deepseek-r1:70b. In contrast, the main disagreeing raters identified overall were the smaller phi-4:14b model and, interestingly, the large open-source llama-4:128x17b.

On \texttt{SummativeChat}, \textit{Anthony2004} resulted a particularly low agreement between LLM-raters due to a contextual misalignment between the schema’s original setting and the analysed data. While \textit{Anthony2004} originally defined the \textit{interface} category for questions regarding the learning platform (e.g., \textit{can I change my answers after I put them in?''} \cite{anthony2004student}), the term in \texttt{SummativeChat} mainly referred to programming \textit{interfaces} used for the task. 
For example, messages such as \textit{``Can you show me the backend API for this wrapper''} was labelled as \textit{interface} by all LLM-raters, confusing the reference to an external \textit{interface}. 
% If `interface' is removed from the schema, the inter-LLM-rater agreement of the newly labelled data increases to moderate-to-good agreement (mean Fleiss’ Kappa = 0.534; mean Gwet's AC1 = 0.690) (random subset of 1000 questions with 7 LLM-raters on \texttt{SummativeChat}).
In contrast, the ambiguity was minimal in \texttt{FormativeChat} as the task was embedded within the learning platform; thus, involving no other \textit{interface}. Therefore, platform-related questions (e.g., \textit{``whats in part g?''} or \textit{``could I switch to 16.1d please?''}) correctly aligned with the original schema, resulting in more consistent labelling.

% \noindent\textbf{Internal Consistency of LLM-Raters: } 
\subsubsection{Internal Consistency of LLM-Raters: } 
To assess whether the order of class definitions within a prompt influenced the classification, we compared LLM-rater agreements using the original schemas against a shuffled version. The  Wilcoxon Signed-Rank Test indicated that the agreement between these versions was statistically significant ($p < .001$), suggesting that LLM-raters were not dependent on the specific order of labels presented in their system prompt. Hence, we utilised the original schema for all analysis in this paper.

% \noindent\textbf{Agreement between LLM and Human Raters: } 
\subsubsection{Agreement between LLM and Human Raters: } 
Integrating the three human experts into the rater pool for \texttt{FormativeChat} resulted in a slight decrease in overall inter-rater agreement, although it remained moderate (mean Fleiss’ Kappa = 0.517; mean Gwet's AC1 = 0.565). The mean agreement between human raters was generally lower than that observed between the LLM-raters on the same subset of student questions: Fleiss’ Kappa = (Human: 0.472, LLM: 0.532) and Gwet's AC1 = (Human: 0.560, LLM: 0.604)). 
% Similar to LLM-raters, human raters also resulted in lower agreement for \textit{Anthony2004} (Fleiss’ Kappa = 0.408; Gwet's AC1 = 0.461) due to the context misalignments of the schema (\Cref{r:anthony_misalignment}).

The differences in agreements between human and LLM-raters suggest that while LLMs are highly consistent with one another, their agreed labels diverged from the interpretations held by human raters.
For instance, the question (a) in \Cref{tab:qualitative_examples} was labelled by majority of LLM-raters as \textit{procedural} in \textit{Cao2021}; but, the human raters vastly disagreed amongst themselves, classifying it as \textit{procedural}, \textit{concept}, and \textit{judgmental}.
We argue that the lower consistency among humans does not necessarily imply inferior correctness compared to the higher consistency of LLMs.
Rather, this variance likely reflects the inherent subjective nuance of student-chatbot interactions and the varying complexity of applying strict classification schemas on such contextual data.

\subsection{(RQ2) What Types of Questions Do Students Ask?}\label{results:proceduralQs}

\begin{figure}[t!]
    \centering
    \includegraphics[width=1\linewidth]{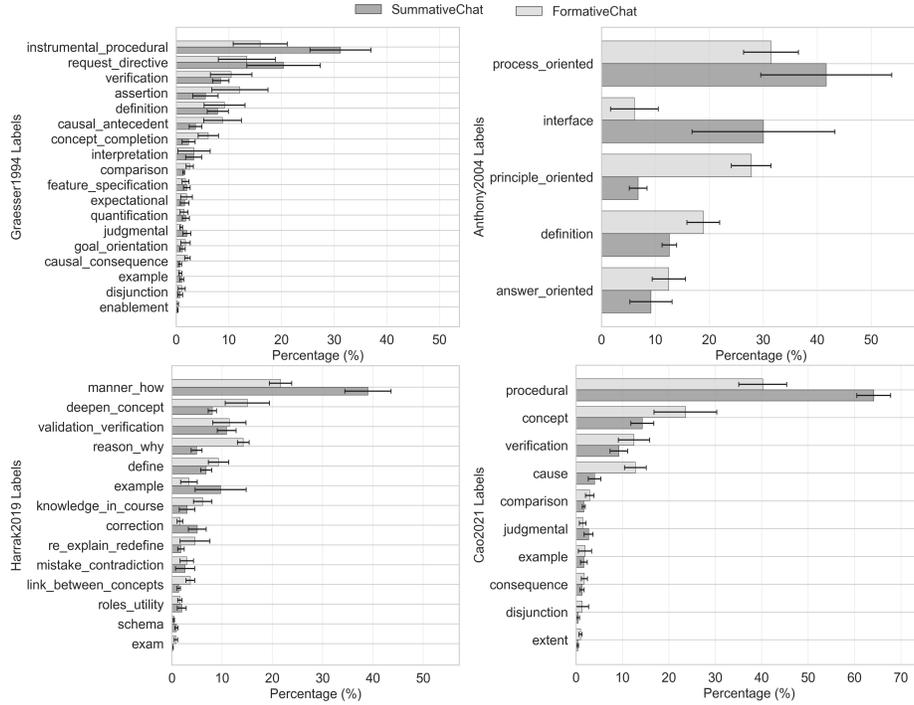}
    
    % \begin{subfigure}
    %     \centering
    %     \includegraphics[width=\linewidth]{misc/QuestionAnalysisPipeline-LLM_Dataset_legend.drawio.png}
    % \end{subfigure}
    % \begin{subfigure}
    %     \centering
    %     \includegraphics[trim={0 0 4.9cm 0},clip, width=0.49\linewidth]{misc/combined_graesser1992_label_distribution.png}
    % \end{subfigure}
    % \begin{subfigure}
    %     \centering
    %     \includegraphics[trim={0 0 4.9cm 0},clip, width=0.49\linewidth]{misc/combined_anthony2004_label_distribution.png}
    % \end{subfigure}
    % \begin{subfigure}
    %     \centering
    %     \includegraphics[trim={0 0 4.9cm 0},clip, width=0.49\linewidth]{misc/combined_harrak2019_label_distribution.png}
    % \end{subfigure}
    % \begin{subfigure}
    %     \centering
    %     \includegraphics[trim={0 0 4.9cm 0},clip, width=0.49\linewidth]{misc/combined_caowang2021_label_distribution.png}
    % \end{subfigure}
    \caption{Distribution of student question types for \texttt{FormativeChat} (light) and \texttt{SummativeChat} (dark), calculated as the mean across 11 LLM-raters for four schemas. Error bars represent standard deviation across models. 
    Results illustrate the prevalence of \textit{procedural} questions across both learning contexts.
    }
    \label{fig:proceduralQsDominate}
\end{figure}

\Cref{fig:proceduralQsDominate} shows that the predominant question across both datasets revolved around asking for the manner in which to proceed or what plan or instrument to use to accomplish a goal (i.e. \textit{instrumental\_procedural}, \textit{process\_oriented}, \textit{manner\_how}, \textit{procedural})(examples in \Cref{tab:exampleRequests}). For clarity, we refer to this type of questions as \textit{procedural} questions in this paper. 

The specific prevalence of \textit{procedural} questions varied significantly by the learning context of the task. 
On average across the four schemas, students in \texttt{SummativeChat} asked \textit{procedural} questions more frequently than those in \texttt{FormativeChat}.
Specifically, in \texttt{SummativeChat}, the prevalence of \textit{procedural} questions was higher by 16.85 percentage points for \textit{Graesser1994}, 8.94 for \textit{Anthony2004}, 19.23 for \textit{Harrak2019}, and 22.94 for \textit{Cao2021}.
% Students in the summative coursework setting (\texttt{SummativeChat}) asked \textit{procedural} questions more frequently than those in the formative self-study setting (\texttt{FormativeChat}) across the schemas. 

\begin{table}[]
    \centering
    \renewcommand{\arraystretch}{1.4}
    \setlength{\tabcolsep}{0.5em}
    \footnotesize
    \caption{\textit{Procedural} questions and \textit{requests} from formative and summative learning contexts labelled by gpt-5.1 in \textit{Graesser1994}.}
    \begin{tabular}{|>{\centering}p{0.155\linewidth}|p{0.39\linewidth}  |p{0.4\linewidth}|} 
        \hline
          \centering\textbf{Label}&\centering\textbf{\texttt{FormativeChat}} & \centering\textbf{\texttt{SummativeChat}} \arraybackslash \\
        \hline
         
          \multirow{4}{*}{\parbox{\linewidth}{\textit{instrumental\_ procedural}}}&
         \textit{``what do i do''}\newline
         \textit{``how do i know that the particle is at y=0''}\newline
         \textit{``okay done that, but how to deal with N having no dimensions?''} 
         & 
        \textit{``How do i run main.py''}\newline
        \textit{``how would i append to the beginning of a string''}\newline
        \textit{``[code snippet] What am I setting this [to] does it use the inputsize?''}
        \\ \hline
        
         \multirow{5}{*}{\parbox{\linewidth}{\textit{request\_{\newline}directive}}}&
        \textit{``help with d) :(''}\newline
        \textit{``can you go through all steps of...''}\newline
        \textit{``im not sure can you just tell me''}\newline
        \textit{``can u [you] do it''}\newline
        \textit{``Can you make a few flash cards...''}\newline
        \textit{``could you give me a summary of the main learning outcomes?''}
        &
        \textit{``[code] explain this portion to me''}\newline
        \textit{``given [data interpretation] can you write [detailed code instructions]''}\newline
        \textit{``[code snippet] What is the error, change this [section of code]''}\newline
        \textit{``Give me the markdown code in the correct format''}
        \\
        \hline
    \end{tabular}
    \label{tab:exampleRequests}
\end{table}

Beyond \textit{procedural} questions, most other common types were not as distinguishable between the two datasets, but some stood out in frequency overall. For example, students asked requests for help (e.g., \Cref{tab:exampleRequests}), conceptual questions regarding course materials or definitions (e.g., from \texttt{SummativeChat}: \textit{``what does read\_csv return in python pandas module''}, \textit{``What does self similarity mean''}, from \texttt{FormativeChat}: \textit{``what does the subscript t mean''}) and verified their own knowledge or validated their guess (e.g., from \texttt{SummativeChat}: \textit{``is an http load balancer a middlebox''}, \textit{``would i be able to use find\_last\_node for [...]''}, from \texttt{FormativeChat}: \textit{``so fbody/ rho = rho g?''}).

\clearpage
\section{Discussion}\label{discussion}
% comparison to schemas, what is the impact, why do I care as a reader (so what?)

\subsection{(RQ1) LLM Raters Agree but Are Restricted by Schemas}\label{d:agreements}

We observed moderate-to-good agreement levels between the LLM-raters, validating the feasibility of labelling student questions at scale. However, some questions highlighted the ambiguities and complexities of student questions, limiting the application of schemas from other contexts to student-LLM conversations.
For instance, question (a) in \Cref{tab:qualitative_examples} split LLM-raters in half between \textit{expectational} and \textit{causal\_antecedent}.

The friction between the context of the schemas and data is highlighted by the low agreements of the \textit{Graesser1994} and \textit{Anthony2004} schemas across both human and LLM-raters (\Cref{fig:looAgreementComparison}). Qualitative feedback from human raters mentioned that \textit{Graesser1994} had a high granularity that led to high cognitive load during classification. Meanwhile, \textit{Anthony2004} was referred to as \textit{``ambiguous''} especially for messages requesting actions from the chatbot or `interface'. 
% These observations suggest the schemas are not fit for student-chatbot conversations, due to the complexity of the questions and the contrasting learning context (see \Cref{tab:qualitative_examples}).
% These observations suggest the schemas are not fit for student-chatbot conversations, as the schemas derive from different learning contexts and, thus, the questions asked during the unique interaction with LLM chatbots are not directly classifiable into a single category (see examples in \Cref{tab:qualitative_examples}) \cite{koyuturk2025learnerchatbotInteractions}.
% These observations suggest that both the inherent complexity and the original learning context of schemas significantly impact the reliability of labelling for both humans and LLMs, supporting similar challenges from other studies \cite{koyuturk2025learnerchatbotInteractions}.
% martinenghi2024VonneumidasSchema
% {\color{red} POSSIBLE IDEA: Maybe if the human experts have a larger variety than for the LLM-raters then this is a task that can be fully trusted to the AI to do it. }
% LLMs can be seen as the wisdom of the crowds as they learned on accumulated knowledge of subsets of crowds (where the bias is coming from)

% \subsection{Are Student Question Schemas Appropriate for Student-Chatbot Interactions?}
\textit{Cao2021} and \textit{Harrak2019} schemas lack specific labels for general student requests, which are frequent in student-chatbot conversations, as covered by \textit{request\_directive} in \textit{Graesser1994}.
However, \textit{Graesser1994} was still limited and struggled to distinguish between student requests driven by different learning goals. For instance, in \texttt{FormativeChat}, a request ranged from an inquiry for clarification (e.g., \textit{``can you go through all steps of \dots''} from \Cref{tab:exampleRequests}) to an attempt at task delegation (e.g., \textit{``can u [you] do it''} from \Cref{tab:exampleRequests}). Grouping these distinct behaviours into a single category misses the intent behind the student question, making it difficult to assess whether the interaction supports or hinders learning. We discuss this further in \Cref{discussion:proceduralQ}. Hence, more incremental definitions of requests (such as \textit{re\_explain\_redefine} in Harrak2019) provide nuance on the question.

Overall, the semantic complexity in stand-alone student messages challenged both LLM and human raters. 
Composite messages containing multiple distinct utterances (such as \Cref{tab:qualitative_examples} (a)) caused high disagreement between raters due to the classification being limited to only one class per question. This constraint possibly lowered the inter-rater agreements due to the overlap of multiple valid labels within one message. 
In contrast, short messages (like \Cref{tab:qualitative_examples} (b)) also posed a risk of mislabelling, as without the full conversation history known to raters, these utterances could have been ambiguous.
Therefore, analysing each message within the conversation flow may allow a more thorough understanding of the nuance of those messages.

While this study successfully observed a contrast between formative and summative contexts, gathering data from a wider range of educational settings -- such as other academic domains and years of study -- would allow for a wider spectrum of student-LLM chatbot interactions during self-study. 

% Another missing label would cover statements regarding student's confusion or lack on knowledge (\textit{assertion}), which would allow to distinguish between student stating their confusion from guessing. 
% Without such distinctions between labels, valuable nuance about the student while conversing with the chatbot is lost. 

% \subsection{What Kind of Questions Do Students Ask in Formative versus Summative Learning Contexts?}

\subsection{(RQ2) What Are Procedural Questions?} \label{discussion:proceduralQ}

% Examples of procedural questions from both datasets:
% \begin{itemize}
%     \item FormativeChat: Do you include the parameter on the left hand side of the equal sign in the number of parameters n you count?, For the next part of the question, how do you identify a parameter being independent to another parameter?
%     \item SummativeChat: how do i check for duplicate rows (same unique ID) in the datasets, how do i save a dataframe to a new csv file and a new json file
% \end{itemize}

Student questions were classified as predominantly \textit{procedural} in student-chatbot conversations, raising the question of how we define procedural questions.
Graesser and Person \cite{graesser1994question} originally classified \textit{procedural} questions as ``deep'' because, in human tutoring, asking ``how'' usually implied high cognitive demand regarding the reasoning behind a problem-solving process. Similarly, Ammari et al. \cite{ammari2025students} argued that practical, goal-directed interactions with LLM chatbots could still reflect deep cognitive engagement. However, findings from the \texttt{SummativeChat} study \cite{mcnichols2025studychat} found that students frequently used chatbots to seek direct, high-level coding solutions in the form of requests for task delegation.
We observed such \textit{procedural} questions and requests involving low cognitive load in both datasets, for instance requests for specific tools and formulas or iterative requests for procedural steps. 

In \texttt{FormativeChat}, questions such as \textit{``im not sure can you just tell me''} or \textit{``can u [you] do it''} (from \Cref{tab:exampleRequests}) may signal low cognitive engagement with the materials and even attempts to offload cognitive effort onto the chatbot by delegating struggles students were intended to overcome.
Such cognitive offloading is an instance of `inversion effects' of AI integration, where over-reliance on the tool risks leading to reduced cognitive engagement \cite{bauer2025ISAR,zhai2024LLMoverreliance}.
However, questions like \textit{``Can you make a few flash cards \dots''}, or `\textit{`could you give me a summary of the main learning outcomes?''} (from \Cref{tab:exampleRequests}) seem to have sought help to further assist the student's understanding rather than bypassing struggles.
Therefore, analysing the nature of \textit{procedural} questions and requests may bring further insight on whether they represent deep reflection on a process or shallow, process-oriented `tool search'.

\section{Conclusions}

This paper investigated applying classification schemas to student questions during conversations with LLM-driven chatbots and proposed a scalable methodology for using LLMs as raters alongside human raters.

Our analysis highlighted that the LLM-raters tended to agree when identifying the types of questions students asked. However, they were limited by current classification schemas, as none of the evaluated schemas could comprehensively capture the unique dynamics of student-LLM chatbot conversations, especially when students ask ambiguous or nuanced questions.
While LLM-raters tended to agree, human raters achieved only moderate agreement.
The higher disagreement among humans raises the question of whether human `ground truth' provides a more valid interpretation of student questions than LLM-raters.

We found that, in both formative and summative contexts, students predominantly ask LLM chatbots procedural questions. This cannot be taken at face value, as it raises important educational concerns, including the distinction between productive requests for help and attempts at cognitive offloading. 

In the future, the development of classification schemas specifically tailored to student-LLM chatbot conversations and other multi-disciplinary methods -- such as conversation analysis -- could be adapted and applied to mitigate the analysis of ambiguous and nuanced student-chatbot interactions.

% Interpreting requests and procedural questions must be through the educational lens of the students. 
%Such frameworks are essential for accurately identifying use patterns that involve risks like cognitive offloading and for designing the next generation of context-aware, pedagogically-aligned learning assistants.

% \input{main/5_limitations}


\bibliographystyle{splncs04}
\bibliography{bib/ref}

@book{Vygotsky1978ZPD,
	title        = {Mind in Society: The Development of Higher Psychological Processes},
	author       = {Vygotsky, L S},
	year         = 1978,
	publisher    = {Harvard University Press},
	address      = {Cambridge, MA}
}

@inbook{VanLehn1988impasse,
	title        = {Toward a Theory of Impasse-Driven Learning},
	author       = {VanLehn, Kurt},
	year         = 1988,
	booktitle    = {Learning Issues for Intelligent Tutoring Systems},
	publisher    = {Springer US},
	address      = {New York, NY},
	pages        = {19--41},
	doi          = {10.1007/978-1-4684-6350-7_2},
	isbn         = {978-1-4684-6350-7},
	editor       = {Mandl, Heinz and Lesgold, Alan}
}

@incollection{johnson2025lambdafeedback,
	title        = {Formative feedback on engineering self-study: Towards 1 million time per year per cohort},
	author       = {Johnson, P and Fenton, J and Ramsden, P and Chatley, R and Ribera-Vicent, M and Karl, L},
	year         = 2025,
	publisher    = {2025 {IEEE} Global Engineering Education Conference ({EDUCON})}
}

@inproceedings{neagu2025LLMsFeedback,
	title        = {Chatbots for Dialogic Feedback during Self-Study: The Importance of Contextual Information},
	author       = {Neagu, Alexandra and Johnson, Peter B and Nelson, Rhodri},
	year         = 2025,
	booktitle    = {EERN Symposium}
}

@article{bauer2025ISAR,
	title        = {Looking beyond the hype: Understanding the effects of AI on learning},
	author       = {Bauer, Elisabeth and Greiff, Samuel and Graesser, Arthur C and Scheiter, Katharina and Sailer, Michael},
	year         = 2025,
	journal      = {Educational Psychology Review},
	publisher    = {Springer},
	volume       = 37,
	number       = 2,
	pages        = 45
}

@article{zhai2024LLMoverreliance,
	title        = {The effects of over-reliance on AI dialogue systems on students' cognitive abilities: a systematic review},
	author       = {Zhai, Chunpeng and Wibowo, Santoso and Li, Lily D},
	year         = 2024,
	journal      = {Smart Learning Environments},
	publisher    = {Springer},
	volume       = 11,
	number       = 1,
	pages        = 28,
	doi          = {10.1186/s40561-024-00316-7}
}

@article{wollny2021chatbotsInEduReview,
	title        = {Are We There Yet? - A Systematic Literature Review on Chatbots in Education},
	author       = {Wollny, Sebastian  and Schneider, Jan  and Di Mitri, Daniele  and Weidlich, Joshua  and Rittberger, Marc  and Drachsler, Hendrik},
	year         = 2021,
	journal      = {Frontiers in Artificial Intelligence},
	volume       = {Volume 4 - 2021},
	issn         = {2624-8212}
}

@article{labadze2023ChatbotsInEdu,
	title        = {Role of AI chatbots in education: systematic literature review},
	author       = {Labadze, Lasha and Grigolia, Maya and Machaidze, Lela},
	year         = 2023,
	journal      = {International Journal of Educational Technology in Higher Education},
	volume       = 20,
	number       = 1,
	pages        = 56,
	doi          = {10.1186/s41239-023-00426-1},
	isbn         = {2365-9440},
	date         = {2023/10/31},
	id           = {Labadze2023}
}

@inbook{CasebourneWegerif2024,
	title        = {The Role of AI Language Assistants in Dialogic Education for Collective Intelligence},
	author       = {Casebourne, Imogen and Wegerif, Rupert},
	year         = 2024,
	booktitle    = {Artificial Intelligence in Education: The Intersection of Technology and Pedagogy},
	publisher    = {Springer Nature Switzerland},
	address      = {Cham},
	pages        = {111--125},
	doi          = {10.1007/978-3-031-71232-6_7},
	isbn         = {978-3-031-71232-6},
	editor       = {Ilic, Peter and Casebourne, Imogen and Wegerif, Rupert}
}

@misc{jurenka2024responsibledevelopmentgenerativeai,
	title        = {Towards Responsible Development of Generative AI for Education: An Evaluation-Driven Approach},
	author       = {Jurenka, Irina and Kunesch, Markus and McKee, Kevin R. and Gillick, Daniel and Zhu, Shaojian and Wiltberger, Sara and others},
	year         = 2024,
	url          = {https://arxiv.org/abs/2407.12687},
	eprint       = {2407.12687},
	archiveprefix = {arXiv},
	primaryclass = {cs.CY}
}

@article{graesser1994question,
	title        = {Question asking during tutoring},
	author       = {Graesser, Arthur C and Person, Natalie K},
	year         = 1994,
	journal      = {American educational research journal},
	publisher    = {Sage Publications},
	volume       = 31,
	number       = 1,
	pages        = {104--137}
}

@article{olney2012question,
	title        = {Question generation from concept maps},
	author       = {Olney, Andrew M and Graesser, Arthur C and Person, Natalie K},
	year         = 2012,
	journal      = {Dialogue \& Discourse},
	volume       = 3,
	number       = 2,
	pages        = {75--99}
}

@article{harrak2019student,
	title        = {From student questions to student profiles in a blended learning environment},
	author       = {Harrak, Fatima and Bouchet, Fran{\c{c}}ois and Luengo, Vanda},
	year         = 2019,
	journal      = {Journal of Learning Analytics},
	volume       = 6,
	number       = 1,
	pages        = {54--84}
}

@inproceedings{anthony2004student,
	title        = {Student question-asking patterns in an intelligent algebra tutor},
	author       = {Anthony, Lisa and Corbett, Albert T and Wagner, Angela Z and Stevens, Scott M and Koedinger, Kenneth R},
	year         = 2004,
	booktitle    = {International Conference on Intelligent Tutoring Systems},
	pages        = {455--467},
	organization = {Springer}
}

@misc{caowang2021qgen,
	title        = {Controllable Open-ended Question Generation with A New Question Type Ontology},
	author       = {Shuyang Cao and Lu Wang},
	year         = 2021,
	url          = {https://arxiv.org/abs/2107.00152},
	eprint       = {2107.00152},
	archiveprefix = {arXiv},
	primaryclass = {cs.CL}
}

@article{mcnichols2025studychat,
	title        = {The studychat dataset: Student dialogues with chatgpt in an artificial intelligence course},
	author       = {McNichols, Hunter and Ikram, Fareya and Lan, Andrew},
	year         = 2025,
	journal      = {arXiv preprint arXiv:2503.07928}
}

@misc{maiti2024studentsinteractllmpoweredvirtual,
	title        = {How Do Students Interact with an LLM-powered Virtual Teaching Assistant in Different Educational Settings?},
	author       = {Pratyusha Maiti and Ashok K. Goel},
	year         = 2024,
	url          = {https://arxiv.org/abs/2407.17429},
	eprint       = {2407.17429},
	archiveprefix = {arXiv},
	primaryclass = {cs.CY}
}

@misc{ammari2025students,
	title        = {How Students (Really) Use ChatGPT: Uncovering Experiences Among Undergraduate Students},
	author       = {Tawfiq Ammari and Meilun Chen and S M Mehedi Zaman and Kiran Garimella},
	year         = 2025,
	url          = {https://arxiv.org/abs/2505.24126},
	eprint       = {2505.24126},
	archiveprefix = {arXiv},
	primaryclass = {cs.HC}
}

@inproceedings{wang2023bertLLMtalkMoves,
	title        = {Can chatgpt detect student talk moves in classroom discourse? a preliminary comparison with bert},
	author       = {Wang, Deliang and Shan, Dapeng and Zheng, Yaqian and Guo, Kai and Chen, Gaowei and Lu, Yu},
	year         = 2023,
	booktitle    = {Proceedings of the 16th international conference on educational data mining},
	pages        = {515--519}
}

@article{alfaraby2024BertLLMsQclass,
	title        = {Analysis of llms for educational question classification and generation},
	author       = {Said {Al Faraby} and Ade Romadhony and  Adiwijaya},
	year         = 2024,
	journal      = {Computers and Education: Artificial Intelligence},
	publisher    = {Elsevier},
	volume       = 7,
	pages        = 100298
}

@article{gani2025tfidfLLMsQclass,
	title        = {Towards enhanced assessment question classification: a study using machine learning, deep learning, and generative AI},
	author       = {Gani, Mohammed Osman and Ayyasamy, Ramesh Kumar and Alhashmi, Saadat M and Alam, Khondaker Sajid and Sangodiah, Anbuselvan and Khaleduzzman, Khondaker and others},
	year         = 2025,
	journal      = {Connection Science},
	publisher    = {Taylor \& Francis},
	volume       = 37,
	number       = 1,
	pages        = 2445249
}

@article{zheng2023judgingLLMjudge,
	title        = {Judging llm-as-a-judge with mt-bench and chatbot arena},
	author       = {Zheng, Lianmin and Chiang, Wei-Lin and Sheng, Ying and Zhuang, Siyuan and Wu, Zhanghao and Zhuang, Yonghao and others},
	year         = 2023,
	journal      = {Advances in neural information processing systems},
	volume       = 36,
	pages        = {46595--46623}
}

@inproceedings{bavaresco2025llmsAsJudge,
	title        = {Llms instead of human judges? a large scale empirical study across 20 nlp evaluation tasks},
	author       = {Bavaresco, Anna and Bernardi, Raffaella and Bertolazzi, Leonardo and Elliott, Desmond and Fern{\'a}ndez, Raquel and Gatt, Albert and others},
	year         = 2025,
	booktitle    = {Proceedings of the 63rd Annual Meeting of the Association for Computational Linguistics (Volume 2: Short Papers)},
	pages        = {238--255}
}

@inproceedings{koyuturk2025learnerchatbotInteractions,
	title        = {Understanding learner-LLM Chatbot interactions and the impact of prompting guidelines},
	author       = {Koyuturk, Cansu and Theophilou, Emily and Patania, Sabrina and Donabauer, Gregor and Martinenghi, Andrea and Antico, Chiara and others},
	year         = 2025,
	booktitle    = {International Conference on Artificial Intelligence in Education},
	pages        = {364--377},
	organization = {Springer}
}

@article{Ma2025promptingHowTo,
	title        = {What Should We Engineer in Prompts? Training Humans in Requirement-Driven LLM Use},
	author       = {Ma, Qianou and Peng, Weirui and Yang, Chenyang and Shen, Hua and Koedinger, Ken and Wu, Tongshuang},
	year         = 2025,
	month        = aug,
	journal      = {ACM Transactions on Computer-Human Interaction},
	publisher    = {Association for Computing Machinery (ACM)},
	volume       = 32,
	number       = 4,
	pages        = {1–27},
	doi          = {10.1145/3731756},
	issn         = {1557-7325}
}

@inproceedings{zamfirescu2023nonexpertPromptingChal,
	title        = {Why Johnny Can't Prompt: How Non-AI Experts Try (and Fail) to Design LLM Prompts},
	author       = {Zamfirescu-Pereira, J.D. and Wong, Richmond Y. and Hartmann, Bjoern and Yang, Qian},
	year         = 2023,
	booktitle    = {Proceedings of CHI Conference on Human Factors in Computing Systems},
	location     = {Hamburg, Germany},
	publisher    = {Association for Computing Machinery},
	doi          = {10.1145/3544548.3581388},
	isbn         = 9781450394215,
	articleno    = 437,
	numpages     = 21
}

@article{cohen1960kappa,
	title        = {A coefficient of agreement for nominal scales},
	author       = {Cohen, Jacob},
	year         = 1960,
	journal      = {Educational and psychological measurement},
	publisher    = {Sage Publications Sage CA: Thousand Oaks, CA},
	volume       = 20,
	number       = 1,
	pages        = {37--46}
}

@article{fleiss1971kappa,
	title        = {Measuring nominal scale agreement among many raters.},
	author       = {Fleiss, Joseph L},
	year         = 1971,
	journal      = {Psychological bulletin},
	publisher    = {American Psychological Association},
	volume       = 76,
	number       = 5,
	pages        = 378
}

@article{derksen2024kappaparadox,
	title        = {The Kappa paradox explained},
	author       = {Derksen, Bastiaan M and Bruinsma, Wendy and Goslings, Johan Carel and Schep, Niels WL},
	year         = 2024,
	journal      = {The Journal of Hand Surgery},
	publisher    = {Elsevier},
	volume       = 49,
	number       = 5,
	pages        = {482--485}
}

@article{landis1977measurementFleiss,
	title        = {The measurement of observer agreement for categorical data},
	author       = {Landis, J Richard and Koch, Gary G},
	year         = 1977,
	journal      = {biometrics},
	publisher    = {JSTOR},
	pages        = {159--174}
}

@article{Gwet2008,
	title        = {Computing inter-rater reliability and its variance in the presence of high agreement},
	author       = {Gwet, Kilem Li},
	year         = 2008,
	journal      = {British Journal of Mathematical and Statistical Psychology},
	volume       = 61,
	number       = 1,
	pages        = {29--48},
	doi          = {10.1348/000711006X126600}
}

@article{cicchetti2001methodologicalGwetAC1,
	title        = {Methodological commentary the precision of reliability and validity estimates re-visited: distinguishing between clinical and statistical significance of sample size requirements},
	author       = {Cicchetti, Domenic V},
	year         = 2001,
	journal      = {Journal of Clinical and Experimental Neuropsychology},
	publisher    = {Taylor \& Francis},
	volume       = 23,
	number       = 5,
	pages        = {695--700}
}

@book{gwet2014handbook,
	title        = {Handbook of Inter-Rater Reliability: The Definitive Guide to Measuring the Extent of Agreement Among Raters},
	author       = {Gwet, Kilem L},
	year         = 2014,
	publisher    = {Advanced Analytics, LLC},
	note         = {See Chapter on Variance Estimation and Jackknife method},
	edition      = {4th}
}

\end{document}